\documentclass[
 prc,
 preprint,
 showpacs,amsmath,amssymb,aps,floatfix,
 superscriptaddress]{revtex4}
 \usepackage{graphicx} 
 \usepackage{dcolumn}  
 \usepackage{bm}       
 \usepackage{amssymb}
 \usepackage{amsmath}

\begin{document}
\title{
Surface-peaked medium effects in the interaction of nucleons with finite nuclei
}
\author{F. J. Aguayo}
\email{faguayo@dfi.uchile.cl}
\affiliation{
Department of Physics - FCFM, University of Chile \\
Av. Blanco Encalada 2008, Santiago, Chile} 

\author{H. F. Arellano}
\email{arellano@dfi.uchile.cl}
\homepage{http://www.omp-online.cl}
\affiliation{
Department of Physics - FCFM, University of Chile \\
Av. Blanco Encalada 2008, Santiago, Chile} 

\date{\today}

\begin{abstract}
We investigate the asymptotic separation of the optical model
potential for nucleon-nucleus scattering in momentum space,
where the potential is split into a medium-independent term 
and another depending exclusively on the gradient of the 
density-dependent $g$ matrix.
This decomposition confines the medium sensitivity of the nucleon-nucleus 
coupling to the surface of the nucleus.
We examine this feature in the context of proton-nucleus 
scattering at beam energies between 30 and 100 MeV and find
that the \emph{pn} coupling accounts for most of this sensitivity.
Additionally, based on this general structure of the optical potential
we are able to treat both, the medium dependence of the
effective interaction and the full mixed density as described
by single-particle shell models. The calculated scattering observables
agree within 10\% with those obtained by Arellano, Brieva and Love
in their momentum-space $g$-folding approach.

\end{abstract}

\maketitle

\section{Introduction}

Microscopic optical model potentials for nucleon-nucleus 
(\emph{NA}) scattering are usually expressed as
the convolution of a two-body effective interaction
with the target ground-state mixed density.
Their realization becomes feasible with the use of 
Brueckner-Bethe-Goldstone $g$ matrices and resorting to 
simplifying assumptions in their coordinate and/or
momentum dependence.
Thus, nuclear medium effects are disclosed by means of 
volume integrals of density-dependent interactions throughout 
the nucleus\cite{Amo00,Ray92}.

In this article we explore in more detail a recent finding 
suggesting that intrinsic nuclear medium effects, namely those
arising from the dependence of the $g$ matrix on the density,
are dominantly localized in the nuclear surface, \emph{i.e.}
regions where the gradient of $g$ is strongest \cite{Are07a}. 
This result is a consequence of a close examination of
the momentum- and coordinate-space structure of a two-body effective 
interaction spherically symmetric in its mean coordinate.
It is demonstrated quite generally that two-body interactions 
can be expressed as a non-trivial sum of 
a medium-independent term and another which
is functionally --and exclusively-- proportional
to the gradient of the reduced \emph{in-medium} interaction.
As a result, the optical potential in 
momentum space becomes the sum of medium-free 
(free $t$ matrix) and medium-dependent ($g$ matrix) contributions, 
the latter depending exclusively on the variations of the effective 
interaction with respect to the mean coordinate.
This feature yields an enhancement of the intrinsic medium effects
in the nuclear surface and suppression in the saturated volume.
We have investigated the manifestation of this selectivity in
\emph{NA} scattering at proton energies below 100 MeV, 
identifying major sensitivity in the \emph{pn} couplings. 

Current trends in nuclear research and applications have resulted in 
the development and construction of novel research facilities around 
the world. 
Such is the case of radioactive ion beam accelerators 
in the US, Europe and Japan \cite{EURISOL,ISOLDE,SPIRAL,RIKEN,MSU}, 
where intense rare isotope beams are produced and collided against 
selected targets. 
As the energy of these beams are projected to reach 500A MeV, 
their scattering from hydrogen targets would result equivalent to 
intermediate-energy proton collisions from an exotic nucleus. 
At energies below 60A MeV the collision would correspond to low-energy 
proton-nucleus scattering, a regime where the inclusion of medium 
effects in the effective interaction is known to be important. 
These unique facilities will expand our access to the neutron
drip line from the region roughly below carbon isotopes to nuclei
as heavy as $^{52}$S, the most neutron-rich nucleus.
Just before the neutron drip line is reached, neutrons occupy weakly 
bound states spread well away from the bulk of the nucleus.
Hence, from the \emph{NA} scattering point of view, 
counting on an approach capable of tracking more selectively 
the various contributions to the optical model potential 
--particularly its surface structure--
could prove useful for studying and interpreting data
from rare isotope beam facilities.
In this article we present and investigate a simplified form 
of the unabridged optical model potential discussed in Ref. \cite{Are07a}, 
paying attention to the surface structure which emerges from the intrinsic
medium effects as implied by the theory.
Additionally, this form allows us to treat explicitly the off-shell 
mixed density, a long standing limitation of the 
microscopic \emph{in-medium} folding approach of Arellano, Brieva 
and Love (ABL) \cite{Are95}, where the Slater approximation is used.

This article is organized as follows. In Section II we outline
the general framework, discuss the general structure of the 
optical model potential in the single scattering approximation, 
introduce the `$\delta g$-folding' approach
and make contact with known approximations.
Additionally, we examine more closely the various contributions to the optical 
potential, their energy and density dependence.
In Section III we present and discuss results from selected applications
of proton elastic scattering at energies between 30 and 100 MeV.
In Section IV we summarize this work and present its main conclusions.

\section{Theoretical framework}
From a broad perspective diverse formal expressions of the
optical model potential for \emph{NA} scattering can be found 
in the literature \cite{Wat53,Fes58,Ker59,Fet65,Vil67}. 
Although they may differ in the way contact is made with the bare 
\emph{NN} potential, they all take the form of 
a ground-state expectation value of a generalized two-body interaction.
Thus, a general representation of the optical model potential for 
collisions of nucleons with kinetic energy $E$ off a composite
target is given by
\begin{equation}
\label{omp}
U({\bm k}',{\bm k};E) = \int d{\bm p}'\;d{\bm p}\; 
\langle {\bm k}' {\bm p}' \mid \hat T(E)\mid {\bm k}\;{\bm p}\;\rangle_{\cal A}
\; \hat\rho({\bm p}',{\bm p})\;,
\end{equation}
where the subscript ${\cal A}$ indicates antisymmetrization.
In general, $\hat T$ contains information about the discrete 
spectrum of the many-body system.
In this expression $\hat\rho({\bm p}',{\bm p})$ represents the
one-body mixed density corresponding to the ground-state of the target.
A comprehensive evaluation of the optical potential considering the
full $\hat T$ matrix would require the solution of the $(A+1)$-body 
system, a formidable challenge.
This difficulty is circumvented by decoupling the two-body effective 
interaction from the ground-state structure, 
a suitable strategy at intermediate and high energies
when the discrete spectrum of the many-body Green's function is away 
from the projectile energy in the continuum.
This allows the use of single-particle models to describe the
target ground state and the Brueckner-Bethe-Goldstone reaction matrix
to represent the effective interaction.

\subsection{Structure of the optical potential}
As expressed in Eq. (\ref{omp}), a central element in the evaluation
of the optical potential is the representation of the
two-body effective interaction. 
Quite generally, regardless of the physics content or particular 
structure conceived for the \emph{NN} interaction, 
the two-body operator $\hat T$ 
in coordinate space requires the specification of four vectors.
This leads to matrix elements of the form
$\langle {\bm r}' {\bm s}'\mid \hat T\mid{\bm r}\; {\bm s}\rangle$,
where ${\bm r}$ and ${\bm s}$ denote the `prior' coordinates of 
the projectile and target nucleon, respectively.
The primed vectors refer to `post' coordinates.
With these definitions, the so called local coordinate ${\bm Z}$
(hereafter referred as the \emph{mean coordinate}) becomes
\[
\bm Z=\textstyle{\frac{1}{4}}(\bm r'+\bm r+\bm s'+\bm s)\;, \nonumber
\]
corresponding to the simple average of the prior and post coordinates of
the interacting pair.

As demonstrated in Ref. \cite{Are07a}, the momentum-space representation
of the $\hat T$ matrix can be cast in terms of a reduced interaction,
$g_{\bm Z}$, in the form
\begin{equation}
\label{wigner}
\langle{\bm k}'{\bm p}'\mid\hat T\mid{\bm k}\;{\bm p}\rangle =
\int\;
\frac{d{\bm Z}}{(2\pi)^3}
\;e^{i{\bm Z}\cdot({\bm Q}-\bm q)}\;
g_{\bm Z}(\bm K_\parallel; {\bm \kappa}',{\bm \kappa})\;.
\end{equation}
In Eq. (\ref{wigner}) 
${\bm Q}={\bm p}'- {\bm p}$, the recoil of the target nucleon;
${\bm q}={\bm k} - {\bm k}'$, the momentum transfer of the projectile;
${\bm K}=({\bm k}+{\bm k}')/2$, the mean momentum of the projectile;
${\bm P}=({\bm p}'+{\bm p})/2$, the mean momentum of the struck nucleon;
and
\begin{subequations}
\begin{eqnarray}
\label{kappa1}
{\bm \kappa}' = \textstyle{\frac{1}{2}} ({\bm k}' -{\bm p}') &=&
\textstyle{\frac{1}{2}}
[{\bm K}-{\bm P} - \textstyle{\frac{1}{2}}({\bm q+\bm Q}) ]\;,\\
\label{kappa2}
{\bm \kappa}  = \textstyle{\frac{1}{2}} ({\bm k}  -{\bm p} ) &=&
\textstyle{\frac{1}{2}}
[{\bm K}-{\bm P} + \textstyle{\frac{1}{2}}({\bm q+\bm Q}) ]\;, 
\end{eqnarray}
\end{subequations}
the post and prior relative momenta, respectively.
Furthermore,
\begin{equation}
\label{kcm}
{\bm K}_\parallel  = {\bm K} + {\bm P} = 
\textstyle{\frac{1}{2}}(\bm k+\bm k'+\bm p+\bm p')\;,
\end{equation}
interpreted as a longitudinal momentum of the interacting 
nucleons \cite{Are07a}.
With these definitions the integrals on 
$({\bm p},{\bm p}')$ in Eq. (\ref{omp}) are accounted for 
by $({\bm P},{\bm Q})$, 
with $d{\bm p}'d{\bm p}=d{\bm Q}\,d{\bm P}$. 

What is interesting about the above representation for $\hat T$ is that it 
prescribes naturally the way the medium dependence of the two-body 
interaction is mapped through the mean coordinate ${\bm Z}$ in 
the reduced interaction. 
To model this dependence we have resorted to infinite 
nuclear matter, a reasonable starting point to incorporate 
leading-order correlations in the nuclear medium.
In this approach, to each coordinate ${\bm Z}$ we associate its 
nuclear isoscalar density $\rho({\bm Z})=[\rho_n(Z)+\rho_p(Z)]/2$, 
therefore a symmetric nuclear matter Brueckner-Bethe-Goldstone 
reaction matrix $g_{\bm Z}$ ($g$ matrix) satisfying 
\begin{equation}
\label{gnm}
\hat g(\Omega)=\hat v+
\hat v\;\frac{\hat Q}{\Omega+i0-\hat h_1-\hat h_2}\;\hat g(\Omega)\;.
\end{equation}
Here $\hat v$ is the bare \emph{NN} potential, 
$\hat h_1$ and $\hat h_2$ the quasi-particle energies at density $\rho$, 
and $\hat Q$ the Pauli blocking operator to suppress occupied intermediate
states.
The corresponding Fermi momentum is given by
\begin{equation}
\label{fermi}
k_F = (3\pi^2\rho)^{1/3}\;.
\end{equation}
In a finite system, namely a system with confined matter distribution, 
we demand that $\lim_{Z\to\infty}\rho({\bm Z})=0$, so that
\begin{equation}
\label{zlimit}
\lim_{Z\to\infty}\,\hat g_{\bm Z}(\Omega)= \hat t(\Omega)\;,
\end{equation}
the free-space $t$ matrix.

In the context of a spherically symmetric matter distribution,
the ${\bm Z}$ integral in Eq. (\ref{wigner}) can be split
in such a way that its asymptotic structure becomes isolated from the
${\bm Z}$-dependent term \cite{Are07a}. Accordingly
\begin{equation}
\label{separation}
\langle{\bm k}'{\bm p}'\mid\hat T\mid{\bm k}\;{\bm p}\rangle
=
\delta({\bm Q}-{\bm q})\,t({\bm K}_\parallel; {\bm\kappa}',{\bm\kappa}) 
{\;-\;} 
\frac{1}{2\pi^2}
\int_{0}^{\infty}Z^3 dZ\;
\Phi_1(Z\mid {\bm Q}-{\bm q}\mid)\;
\frac{\partial g_Z}{\partial Z}\;,
\end{equation}
where the momentum dependence of $\partial g_Z/\partial Z$ 
on ${\bm K}_\parallel$, ${\bm\kappa}'$ and ${\bm\kappa}$ is implicit.
Here $\Phi_1(t)=j_1(t)/t$, with $j_1$ the spherical Bessel function of order 1.
This profile function favors the recoil of 
the struck nucleon around ${\bm Q} \approx {\bm q}$, 
namely ${\bm k}+{\bm p}\approx{\bm k}'+{\bm p}'$. 
We observe that total momentum conservation can only be possible when 
the system exhibits translational invariance, as expressed when
$\partial g_Z/\partial Z=0$.

Upon substitution of $\hat T$ from Eq. (\ref{separation}) into 
Eq. (\ref{omp}) for the optical potential we obtain 
\begin{equation}
\label{u}
U=U_0+U_1\;,
\end{equation}
with
\begin{subequations}
\begin{eqnarray}
\label{kmt}
U_{0}(\bm k',\bm k;E)&=&\int \;d{\bm P}\;\hat\rho(\bm q;\bm P)\;
t({\bm K}_\parallel; {\bm\kappa}',{\bm\kappa}) \;;\\
\label{una}
U_{1}(\bm k',\bm k;E) &=& 
\frac{1}{2\pi^2}
\int d{\bm Q}\;d{\bm P}\;
\hat\rho(\bm Q;\bm P)\; \times \nonumber \\
& & \int_{0}^{\infty}Z^3 dZ\;
\Phi_1(Z|\bm Q-\bm q|)\; \left(-
\frac{\partial g_Z}{\partial Z} \right)
\;.
\end{eqnarray}
\end{subequations}
The first term, $U_{0}$, depends exclusively on the medium-free reduced 
matrix, whereas the second depends on the gradient of the $g$ matrix. 
In these expressions $\hat\rho$ denotes the full mixed-density, 
which in terms of occupied single-particle states $\phi_\alpha$ is given by
\[
\hat\rho(\bm Q;\bm P)\equiv 
\sum_\alpha \hat\rho_\alpha(\bm Q;\bm P) = 
\sum_\alpha \phi_\alpha^\dagger({\bm p}')\,\phi_\alpha({\bm p})\;.
\]

Eq. (\ref{u}) for $U$ represents the most general expression
to be given to the optical model potential when the two-body effective
interaction exhibits spherical symmetry in the mean coordinate ${\bm Z}$.
It summarizes the medium dependence of a general two-body effective
interaction, accounting for all phase-space configurations
allowed by the one-body mixed density.
The interaction is evaluated off-shell, with no assumptions
regarding its local/nonlocal structure.
Furthermore, it involves a sevenfold integral, 
sixfold in momentum space and an
additional integration in coordinate space. 
Thus, its evaluation constitutes a very challenging task even
for nowadays computational capabilities.
In this work we circumvent this difficulty by introducing
a simplifying assumption, 
within the \emph{momentum-conserving approximation},
to be explained in the following subsection.
A thorough assessment of this assumption and its implications
in actual scattering processes may require the
evaluation of the sevenfold optical potential itself.

\subsection{Limit cases and the $\delta g$-folding}
The general form of the optical potential expressed above
leads naturally to the free $t$ matrix and ABL folding approaches. 
For instance, if the effective interaction is taken as the 
(transitionally invariant) free $t$ matrix, then $U_1$ vanishes
and $U$ becomes $U_0$, the free $t$-matrix full-folding
optical model potential applied to intermediate-energy \emph{NA}
scattering \cite{Are89,Are90a}.
In this case the medium effects do not come from the effective
interaction but from the Fermi motion of the struck nucleons, 
as allowed by the spread of the one-body mixed density. 

The evaluation of the optical potential for a spherically 
symmetric system involves a 7-fold integral. 
In order to simplify this, we neglect the dependence 
of $\partial g_Z/\partial Z$ on ${\bm Q}$
by setting ${\bm Q}\to{\bm q}$ in the interaction. 
This change is designated as $g_Z\to g_Z^{(0)}$, so that
\begin{equation}
\label{mca}
U_1\approx 
\frac{1}{2\pi^2}
\sum_\alpha
\int_{0}^{\infty}Z^3 dZ
\int d{\bm P}\;
\Omega_\alpha(\bm q,\bm P;Z)\; 
\left( -\frac{\partial g_Z^{(0)}}{\partial Z}\right) \;,
\end{equation}
with
\[
\Omega_\alpha({\bm q},{\bm P};Z)=\int d{\bm Q}\;
\hat\rho_\alpha(\bm Q;\bm P)\;
\Phi_1(Z|\bm Q-\bm q|)\;.
\]
Noting that ${\bm Q}\to{\bm q}$ expresses momentum conservation 
of the interacting pair in the $g$ matrix
(${\bm k}+{\bm p}={\bm k}'+{\bm p}'$), we find appropriate to
refer to this as momentum-conserving approximation (MCA). 
The appealing feature of this result is that it enables 
a detailed treatment of the full-mixed density as obtained 
from single-particle shell models
while accounting for the medium dependence in the $g$ matrix. 
The following discussions are mainly focused on this structure of the
optical potential, referred hereafter as `$\delta g$-folding'.

As demonstrated in Ref. \cite{Are07a}, the use of the Slater approximation
within the MCA leads to the ABL potential
\begin{equation}
\label{abl}
U_{ABL}=4\pi
\int_{0}^{\infty}{Z}^2 dZ\;j_{0}(qZ)\,\rho(Z)\;\langle g_{Z}^{(0)}\rangle\;,
\end{equation}
where $\langle g_{Z}^{(0)}\rangle$ denotes the Fermi-motion integral
\[
\langle g_{Z}^{(0)}\rangle = 
\int d{\bm P}\;S_F(P;Z)\,g_Z[{\bm K}_\parallel;
\textstyle{\frac{1}{2}}({\bm K}-{\bm P} - {\bm q}),
\textstyle{\frac{1}{2}}({\bm K}-{\bm P} + {\bm q})]\;,
\]
with 
\[
S_F(P;Z)=\frac{1}{\frac{4}{3}\pi \hat k^3(Z)}\;\Theta[\hat k(Z) - P]\;.
\]
This step-function sets bounds for the off-shell sampling of the
$g$ matrix at a distance $Z$ from the center of the nucleus, 
$\mid{\bm P}\mid\leq \hat k(Z)$, with $\hat k(Z) = [3\pi^2\rho(Z)]^{1/3}$.
The above result for $U_{ABL}$ can also be obtained by replacing directly
the two-body effective interaction [c.f. Eq. (\ref{wigner})] 
into Eq. (\ref{omp}) for $U$, applying the MCA  
and representing the mixed density by its Slater form.

All the above forms of the optical potential are nonlocal,
as a consequence of the momentum structure of the
$g$ matrix --solution of the Brueckner-Bethe-Goldstone integral equation--
expressed in terms of the relative momenta ${\bm\kappa}'$ and ${\bm\kappa}$
[c.f. Eqs. (\ref{kappa1},\ref{kappa2})].
The antisymmetrization of the interaction accounts for
additional nonlocalities.
These features have not been duly explained in previous works, 
leaving room for misconceptions.
So it may be worth to sketch them here for clarity. 
To make the illustration simple let us consider the rank-0 (scalar)
antisymmetrized reduced $g$ matrix for total spin $S$ and isospin $T$,
\[
\langle {\bm\kappa}'\mid g^{ST}\mid{\bm\kappa}\rangle_{\cal A}=
g^{ST}({\bm\kappa}',{\bm\kappa}) - (-)^{S+T}g^{ST}({\bm\kappa}',-{\bm\kappa})\;.
\]
Expanding in partial waves
\[
g^{ST}({\bm\kappa}',{\bm\kappa})=\sum_{L=0} g^{ST}_{L}(\kappa',\kappa)\,
P_L(\hat\kappa'\cdot\hat\kappa)\;,
\]
and using the property $P_L(-u)=(-)^LP_L(u)$, 
we can arrange the antisymmetrized $g$ in a single sum,
\[
\langle {\bm\kappa}'\mid g^{ST}\mid{\bm\kappa}\rangle_{\cal A}=
\sum_{L=0}g^{ST}_{L}(\kappa',\kappa)\,[1-(-)^{L+S+T}]\,
P_L(\hat\kappa'\cdot\hat\kappa)\;.
\]
Therefore
\begin{equation}
\label{antisymm}
\langle {\bm\kappa}'\mid g^{ST}\mid{\bm\kappa}\rangle_{\cal A}=
2\sum_{\textrm{Allowed}}g^{ST}_{L}(\kappa',\kappa)\,
P_L(\hat\kappa'\cdot\hat\kappa)\;,
\end{equation}
where the summation considers only those \emph{NN} states allowed by the
Pauli exclusion principle and 
the off-shell matrix elements $g^{ST}_{L}(\kappa',\kappa)$, 
direct solutions to the Brueckner-Bethe-Goldstone equation 
for the corresponding partial wave.
In this fashion we naturally account for the knock-out exchange term.

In the case of local effective interactions \cite{Amo00,Ger84,Ray99},
the off-shell matrix elements $g^{ST}({\bm\kappa}',{\bm\kappa})$
are obtained \emph{via} the Fourier transform $\tilde v$ 
of the local function $v^{ST}(r)$, 
hence
\[
g^{ST}({\bm\kappa}',{\bm\kappa})= \tilde v^{ST}({\bm\kappa}'-{\bm\kappa})\;.
\]
Therefore, the antisymmetrized matrix element reads
\[
\langle {\bm\kappa}'\mid g^{ST}\mid{\bm\kappa}\rangle_{\cal A}=
\tilde v^{ST}({\bm\kappa}'-{\bm\kappa})-
(-)^{S+T}\tilde v^{ST}({\bm\kappa}'+{\bm\kappa})\;,
\]
a well known result. 
Here the knock-out exchange term makes
the antisymmetrized interaction non local. 
If one uses multipole expansions to these Fourier transforms, then
the antisymmetrized $\langle g^{ST}\rangle_{\cal A}$ 
takes the same form as that expressed by Eq. (\ref{antisymm}). 
In this case, however, 
$g_L^{ST}(\kappa',\kappa)=\int_0^\infty r^2 
j_L(\kappa'r) v^{ST}(r)\,j_L(\kappa r)\,dr$.

\subsection{The medium-dependent term}

We examine more closely the structure of $U_1$, particularly
the shape of its integrands.
Since the dependence of $g$ matrix elements on $Z$ is set \emph{via}
the isoscalar density $\rho$, with $\rho = k_F^3/3\pi^2$, then we can write
\[
\frac{\partial g_Z}{\partial Z} = 
\left . \left (\frac{\partial g}{\partial k_F}\right)\right |_{k_F=\hat k(Z)}
\hat k'(Z)\;,
\]
with
\[
\hat k'(Z)=\frac{\hat k(Z)}{3}\,\frac{\partial \ln \rho}{\partial Z}\;.
\]

\begin{figure}[!ht]
\includegraphics[scale=0.8,angle=-00] {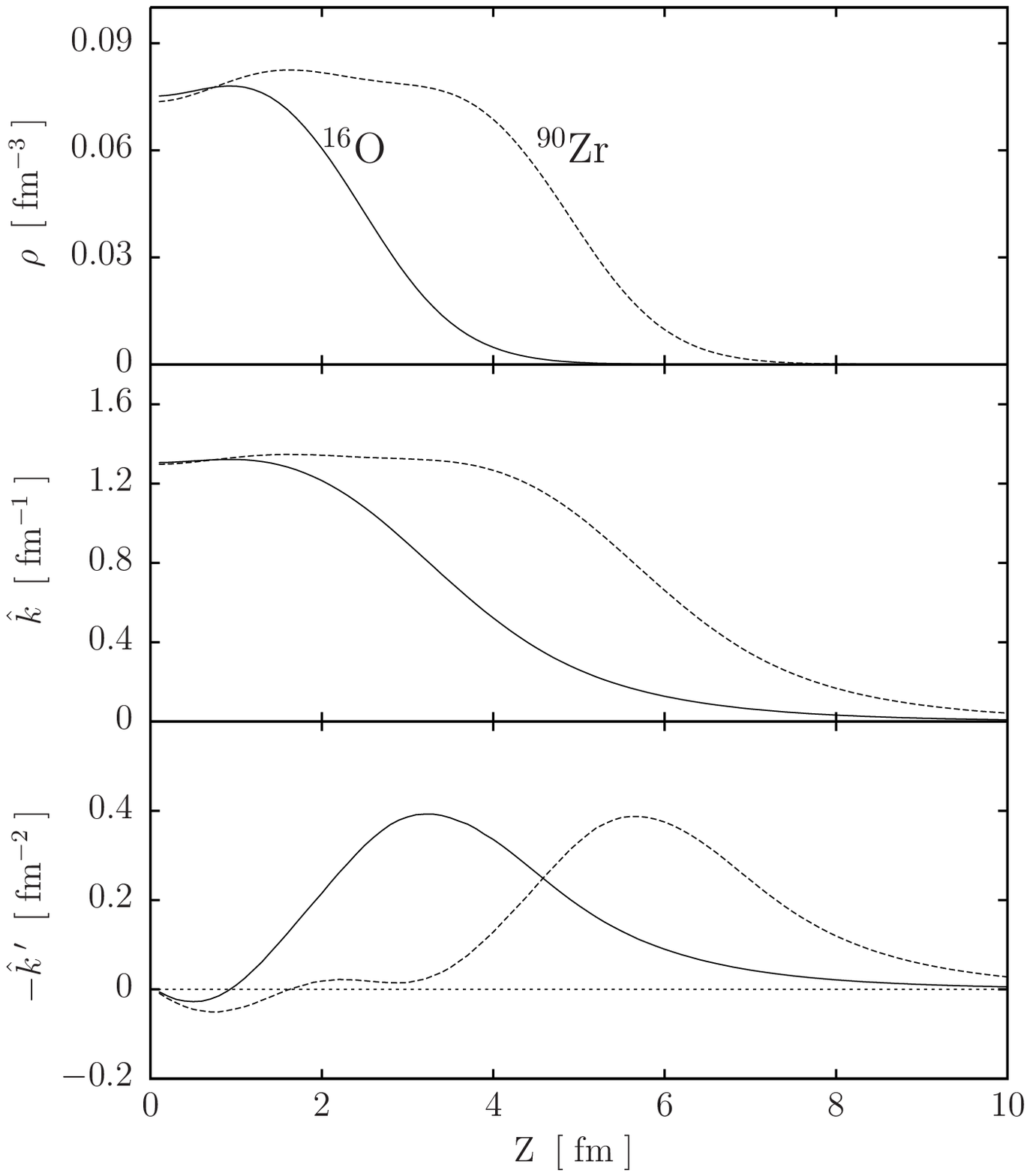}
\medskip
\caption{{\protect\small
\label{mca_dkdz}
     Radial dependence of the density (upper frame), 
     local Fermi momentum (middle frame) and
     its negative gradient (lower frame) for $^{16}$O and $^{90}$Zr.}
        }
\end{figure}

In Fig. (\ref{mca_dkdz}) we plot the radial dependence of the 
density $\rho(Z)$ (upper frame), 
its corresponding local Fermi momentum $\hat k(Z)$ (middle frame)
and the negative radial derivative $-\hat k'(Z)$ (lower frame),
for $^{16}$O (solid curves) and $^{90}$Zr (dashed curves), respectively.
These figures exhibit clear peaks of $-\hat k'(Z)$  near 3 fm and 6 fm,
corresponding in both cases to $\hat k\approx$ 0.6 fm$^{-1}$,
\emph{i.e.} the surface of the nucleus.
We estimate in $\sim$3 fm the width of both peaks, 
limiting the region where the main contributions to $U_1$ should come from.
The strength of these contributions are dictated by the derivative 
$\delta g\equiv\partial g/\partial k_F$, which depends on the energy $E$
of the projectile.

\begin{figure}[!ht]
\includegraphics[scale=0.8,angle=-00] {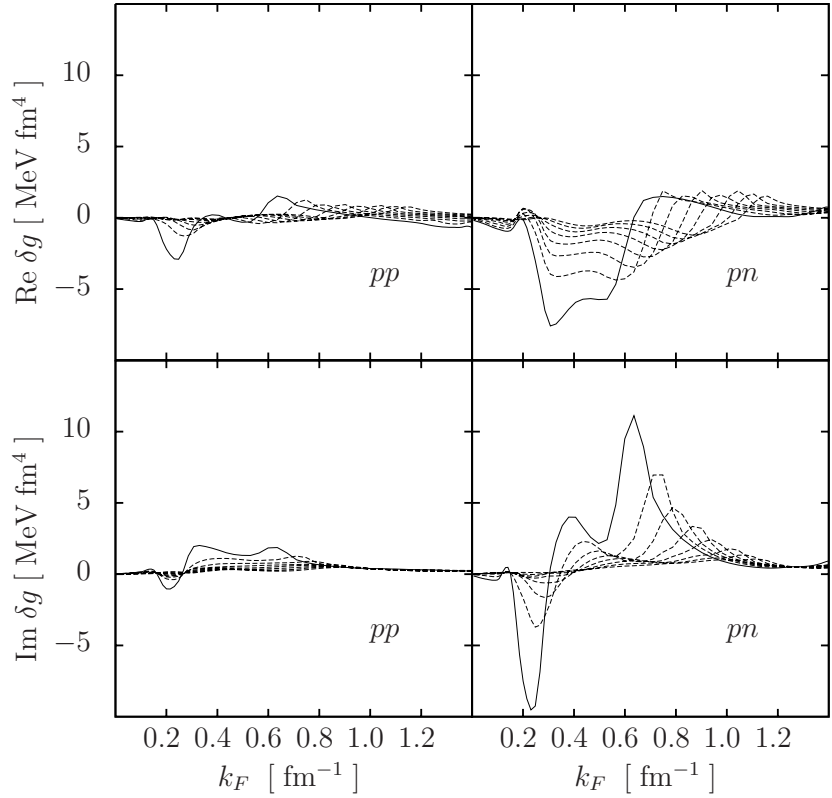}
\medskip
\caption{{\protect\small
\label{mca_dg}
         Real (upper frames) and imaginary (lower frames) 
         components of the forward 
         $\delta g\equiv\partial g/\partial k_F$ amplitudes, in the
         \emph{pp} (left frames) and \emph{pn} (right frames)
         channels, as functions of the Fermi momentum.
         See the text for reference to the curves.
        }
        }
\end{figure}

\begin{figure}[!ht]
\includegraphics[scale=0.8,angle=-00] {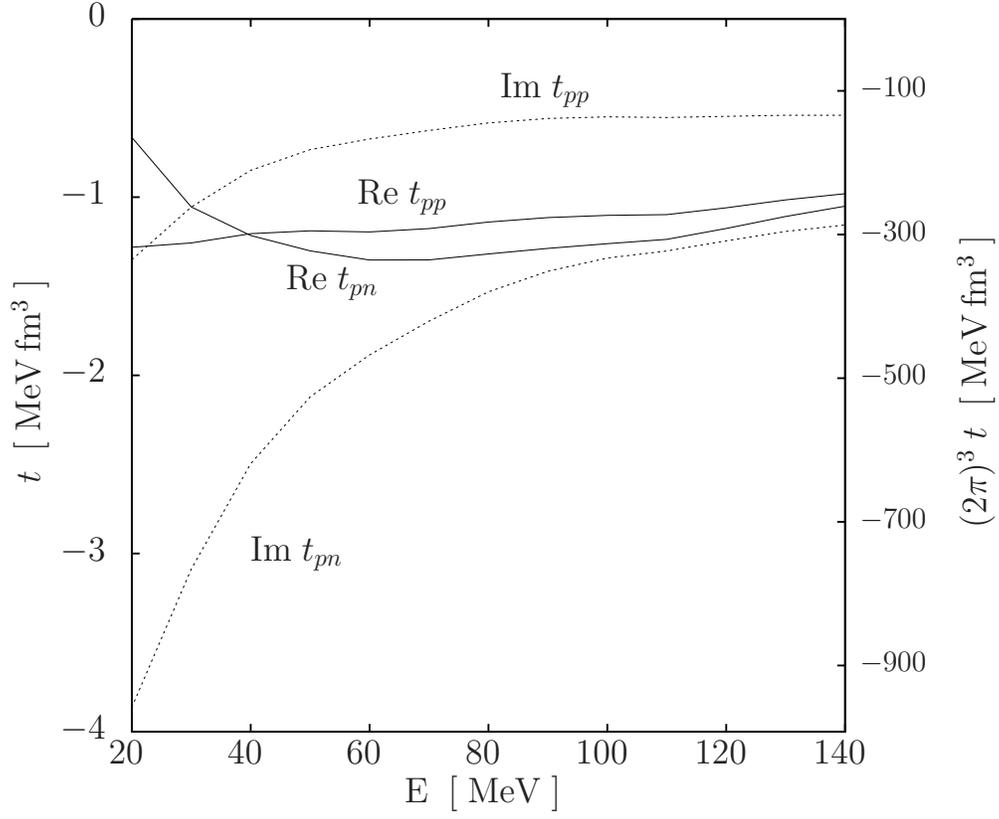}
\medskip
\caption{{\protect\small
\label{mca_tmatrix}
         Forward on-shell free $t$ matrix as function of the nucleon energy.
         The solid and dashed curves correspond to the real and
         imaginary amplitudes, respectively.
        }
        }
\end{figure}

In Fig. (\ref{mca_dg}) we show the partial derivative of the on-shell
$g$ amplitude with respect to the Fermi momentum,
symbolized with $\delta g$.
The real and imaginary components are shown in the upper and lower
frames, respectively. The left frames correspond to the \emph{pp}
channel, whereas the right frames correspond to the \emph{pn} channel.
The curves represent different projectile energies, starting 
at $E=$30 MeV (solid curves) up to 100 MeV in steps of 10 MeV (dashed curves). 
To facilitate their comparison, the same scale is used in all graphs.
By forward (on shell) we mean ${\bm k}'={\bm k}$, with $E=k^2/2m$, 
the nonrelativistic nucleon energy. 
Hence, $\delta g=
\partial g({\bm k};
\textstyle{\frac{1}{2}}{\bm k},
\textstyle{\frac{1}{2}}{\bm k})/\partial k_F$.
The striking feature of this figure is the asymmetry of $\delta g$,
significantly more pronounced in the \emph{pn} than in the \emph{pp}
channel, suggesting more sensitivity to neutron densities 
when protons are used as projectiles. 
Looking at the real part of the \emph{pn} coupling,
the attraction is more pronounced in the region 
0.2 fm$^{-1} \lesssim k_F\lesssim$ 0.6 fm$^{-1}$, 
\emph{i.e.} the nuclear surface, a feature which diminishes with 
increasing energy. 
Regarding the imaginary contribution, 
the nuclear surface contributes with more absorption, 
whereas in the nuclear interior ($kF\gtrsim$ 0.6 fm$^{-1}$) it is weakened.
It is important in this analysis to keep in mind that 
$\delta g$ contributes to $U_1$. 
Instead, the leading-order contribution to the optical potential 
stems from $U_0$, which depends directly on the $t$ matrix. 
To keep this observations in better perspective,
in Fig. (\ref{mca_tmatrix}) we plot the forward (on shell) 
$t$ matrix as function of the nucleon energy $E$.
Here the solid and dashed curves correspond to the real and imaginary 
amplitude, respectively.
In this figure the right-hand-side axis scales to $(2\pi)^3\,t$, 
to facilitate comparison with other conventional normalizations.
We notice here that the absorptive component of the \emph{pn}
coupling exhibits a stronger energy dependence, 
becoming dominant as the energy decreases from $\lesssim$ 80 MeV.
Instead, the real components are relatively constant throughout
the energy range considered.

\begin{figure}[!ht]
\includegraphics[scale=0.8,angle=-00] {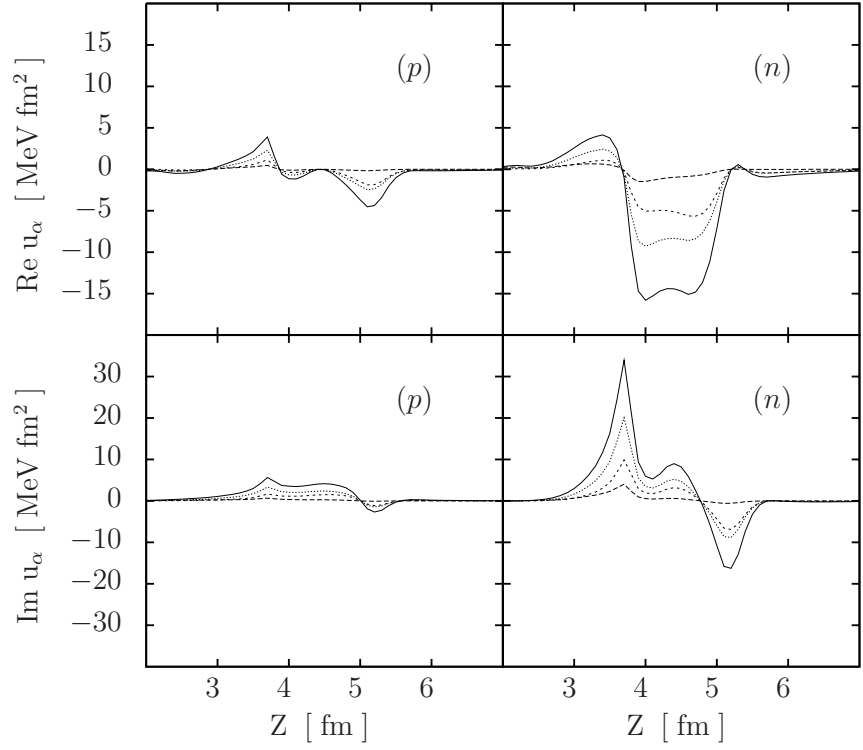}
\medskip
\caption{{\protect\small
\label{mca_ualpha}
         The radial behavior of $u_\alpha$ for the \emph{pp} (left frames)
         and \emph{pn} couplings (right frame).
         Observe that $\sum_\alpha\int u_\alpha^{(p,n)}(Z)\,dZ=U_{1}^{(p,n)}$.
         See text for reference to the curves.
        }
        }
\end{figure}

In order to trace the sources of the contributions to $U_1$ 
and also estimate their importance relative to $U_0$,
we find useful to introduce the density function $u_\alpha$ defined by
\begin{equation}
\label{u_alpha}
u_\alpha(Z)= -\,\frac{Z^3}{2\pi^2} \int\,d{\bm P}\,
\Omega_\alpha({\bm q}=0,{\bm P};Z)\,
\frac{\partial g_Z^{(0)}}{\partial Z}\,,
\end{equation}
to be evaluated on-shell at ${\bm q}=0$ for the 
single-particle shell $\alpha$. 
Its radial integral accounts for the partial contributioin $U_\alpha$,
\[
U_\alpha=\int_0^\infty u_\alpha(Z)\,dZ\;,
\]
with $U_1=\sum_{\alpha}U_\alpha$.
In Fig. (\ref{mca_ualpha}) we plot the real (upper frames)
and imaginary (lower frames) components of $u_\alpha(Z)$ 
for $^{16}$O(\emph{p},\emph{p}) at 30 MeV.
The curves correspond to contributions from the 
1$p_{3/2}$ (dotted curves), 
1$p_{1/2}$ (dashed curves) and 
1$s_{1/2}$ (long-dashed curves) shells,
while the solid curves represent the sum $\sum_\alpha u_\alpha^{(p,n)}$. 
The (\emph{p}) and (\emph{n}) labels symbolize contributions
of the form $\langle\hat\rho_p\,\delta g_{pp}\rangle$,
arising from proton densities, and 
$\langle\hat\rho_n\,\delta g_{pn}\rangle$ due to neutrons, respectively. 
Notice that the scale of \emph{Im} $u_\alpha$ doubles 
that of \emph{Re} $u_\alpha$.

This figure evidences quite neatly surface-peaked structure
stemming from $\delta g$, confined in the region 3-5.5 fm, with 
clear dominance of neutron over proton distributions.
Considering $U_0$ the leading-order contribution,
\emph{Re} $u_\alpha^{(n)}$ enhances the attraction to the projectile (proton).
This can be readily estimated considering its width $\sim$1.5 fm
and depth $\sim$15 MeV fm$^2$. Hence, the area between the curve and
the $Z$ axis yields \emph{Re} $U_1^{(n)} \sim$ -23 MeV fm$^3$.
This is to be compared to
$U_0^{(n)}\approx 8\times$\emph{Re} $t_{pn}\sim$ -10 MeV fm$^3$,
as extracted from Fig. (\ref{mca_tmatrix}).
In contrast, neutron density contributions to \emph{Im} $U_1$ becomes 
considerably weaker due to its near-canceling up-and-down structure 
observed in the lower-right frame, while 
\emph{Im} $U_0^{(n)} \approx 8\times$\emph{Im} $t_{pn}\sim$ -24 MeV fm$^3$.
The extent to which these features become important in collision processes
needs to be assessed by examining scattering observables.
In any case, the pocket shape of \emph{Re} $u_\alpha^{(n)}$ near the
surface indicates a preference to couple the projectile (proton) 
with the $\nu 1p_{1/2}$ and $\nu 1p_{3/2}$ shells, 
favoring $(p,d)$ pickup reactions. 
This feature is consistent with recent findings on pickup effects 
in $p+^{10}$Be elastic scattering near 40A MeV \cite{Kee08}.


\section{Applications}
We investigate proton elastic scattering from $^{16}$O and $^{90}$Zr, 
two relatively well known doubly closed-shell nuclei.
In each case we consider three forms of the optical potential.
First, the $\delta g$-folding approach [c.f. Eq. (\ref{mca})], 
providing arguably the most complete momentum-space description of the 
optical potential to date.
Here, single-particle wavefunctions are used to represent the
one-body full mixed density while a thorough account of the medium dependence
of the antisymmetrized off shell $g$ matrix is given. 
Second, the ABL folding approach, corresponding to a simplified
representation of the mixed density in terms of its Slater approximation.
This approach has been extensively discussed in Refs. \cite{Are95,Are02}.
Lastly, the free $t$-matrix full-folding optical potentials ($t$-folding), 
where the full mixed density is used as in the early calculations
\cite{Are89,Are90a}.

The calculations presented here are based on the Paris 
\emph{NN} potential \cite{Lac80}.
We have investigated other \emph{NN} potentials and found no significant
differences with the results reported here.
The corresponding $g$ matrix was calculated off shell ($J\leq 7$) 
at 30 values of the Fermi momentum, ranging from 0 up to 1.6 fm$^{-1}$.
This thin mesh is no longer necessary after various tests
of convergence were performed; the use of 16 Fermi momenta yields
equally reliable results.
The needed selfconsistent nuclear-matter fields were computed 
prior to all runs.

To evaluate $U_1$ given by Eq. (\ref{mca}) we carry out the
${\bm P}$ and ${\bm Q}$ integrals using Gauss-Laguerre quadrature
at 25 radial mesh points.
The $Z$ integration is performed using a uniform mesh with steps of 0.1 fm.
As stated earlier, the calculated optical model potentials reported 
here are nonlocal operators, treated as such throughout. 
The scattering observables are obtained solving the Schr\"odinger
equation with the nonlocal coupling in the presence of
the Coulomb term. See Ref. \cite{Are07b} for more details.

\subsection{ $p+^{90}$Zr scattering}

\begin{figure}[!ht]
\includegraphics[scale=0.8,angle=-00] {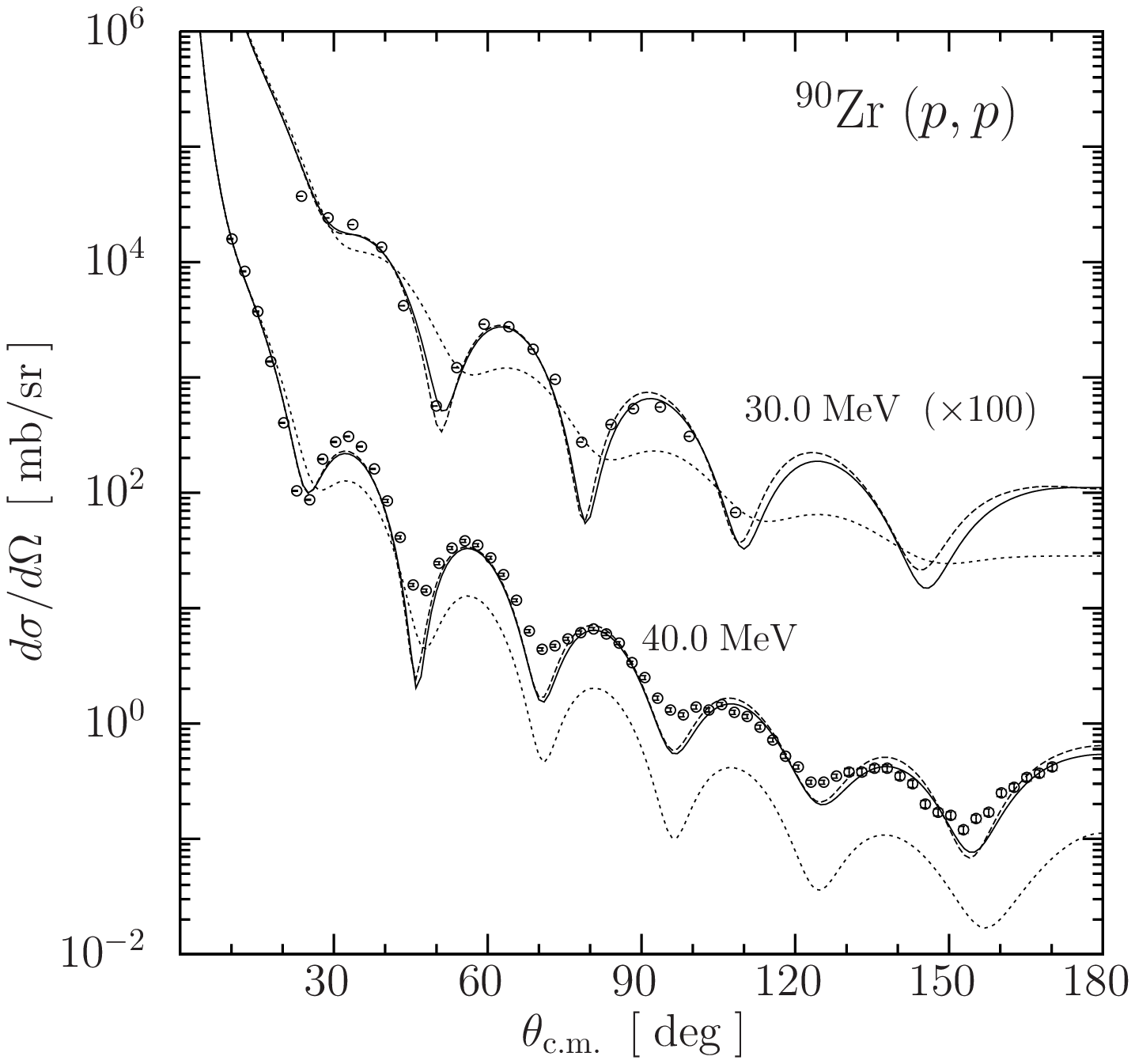}
\medskip
\caption{{\protect\small
\label{mca_zr2a}
   The measured and calculated differential cross section for
$^{90}$Zr(\emph{p},\emph{p}) at 30.4 and 40 MeV.
The data are from Refs. \cite{Swi77} and \cite{Blu66}, respectively.
See text for reference to the curves.
        }
        }
\end{figure}

In Fig. (\ref{mca_zr2a}) we present the measured
and calculated differential cross section, as a function of the
center-of-mass scattering angle $\theta_{c.m.}$, for 
$^{90}$Zr(\emph{p},\emph{p}) at 30.4 and 40 MeV.
The data are from Refs. \cite{Swi77} and \cite{Blu66}, respectively.
The solid curves correspond to $\delta g$-folding, 
the dashed curves to the ABL approach and the dotted curves to the $t$-folding.
The one-body mixed density is constructed using single-particle 
wavefunctions based on Hartree-Fock calculations \cite{Neg70}.

The calculated cross sections based on the $\delta g$-folding follow
reasonably well the diffractive pattern exhibited by the measurements.
The maxima are in phase with the data, although the diffractive 
minima tend to be more pronounced. 
Additionally, we find that the ABL approach (dashed curves) 
follows very closely the $\delta g$-folding (solid curves). 
So, within the MCA, 
the ABL folding approach represents reasonably well the $\delta g$-folding. 
The difference lies in the computational time needed for their evaluation, 
being the $\delta g$-folding more time demanding 
(by nearly a factor 150) than its ABL counterpart.

Considering the results based on the free $t$ matrix (dotted curves), 
they clearly lack the structure exhibited by the data. 
In particular, at the two energies considered here,
the first diffractive minima occur at greater angles 
than those shown by the data, suggesting a smaller size nucleus. 
Thus, the medium effects accounted for by $U_1$, and located
mainly in the nuclear surface, do account for some of the hadronic size 
of the nucleus.

\begin{figure}[!ht]
\includegraphics[scale=0.8,angle=-00] {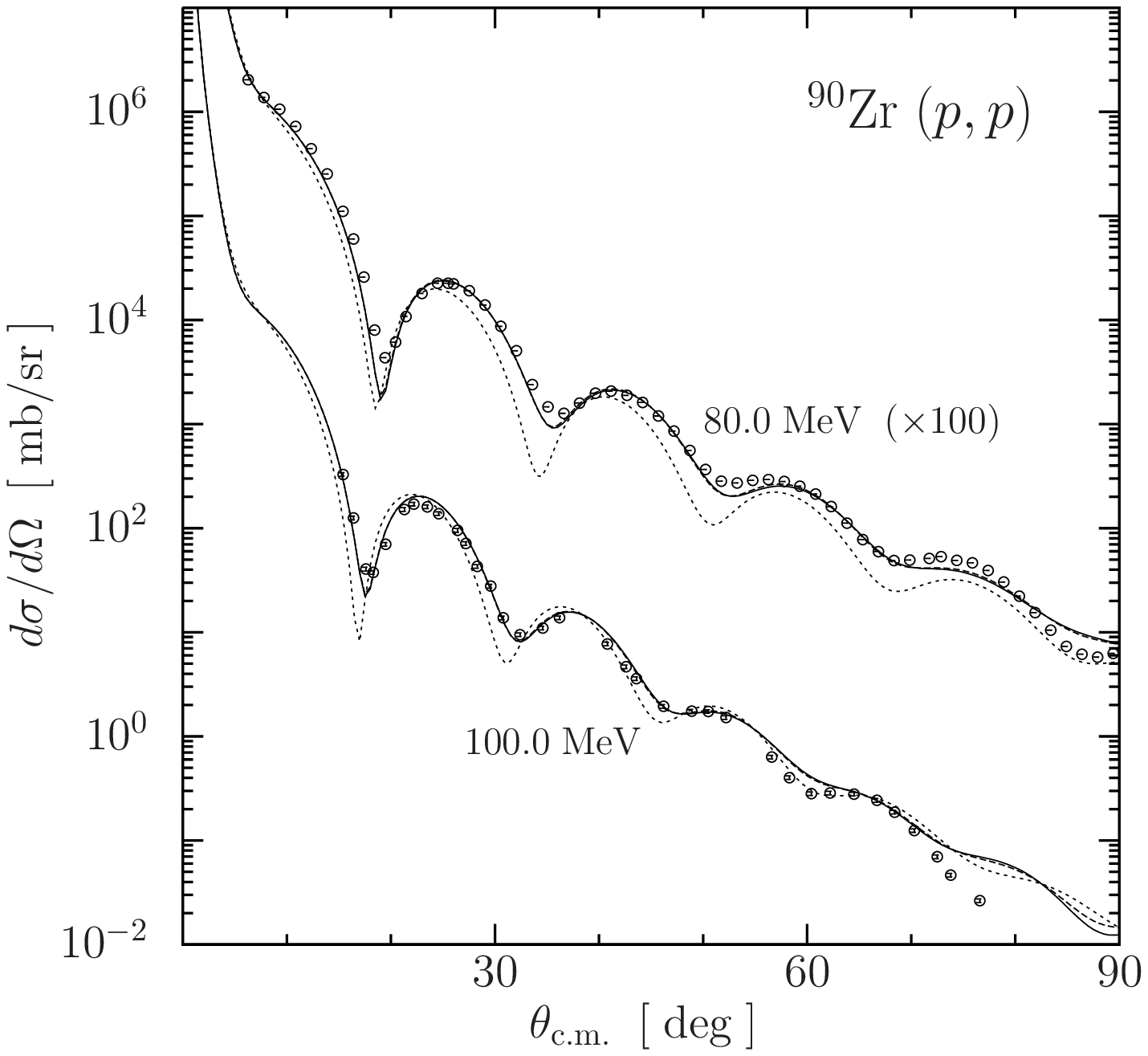}
\medskip
\caption{{\protect\small
\label{mca_zr2b}
   The measured and calculated differential cross section for
$^{90}$Zr(\emph{p},\emph{p}) at 80 and 100 MeV.
The data are from Refs. \cite{Nad81} and \cite{Kwi78}, respectively.
See text for reference to the curves.
        }
        }
\end{figure}

As the energy increases it is expected that the medium effects become
less relevant in the scattering process. This feature 
is clearly observed in Fig. (\ref{mca_zr2b}), where we plot the
measured and calculated differential cross section for
$^{90}$Zr(\emph{p},\emph{p}) at 80 and 100 MeV.
The data are from Refs. \cite{Nad81} and \cite{Kwi78}, respectively.
The curves follow the same convention as in Fig, (\ref{mca_zr2a}).
The agreement of the $\delta g$-folding with the data is remarkable 
throughout the whole range of the measurements.
Additionally, we verify that the difference between the $t$-folding 
results and those from $\delta g$- or ABL-approach have 
diminished considerably relative the previous applications.

\subsection{ $p+^{16}$O scattering}

\begin{figure}[!ht]
\includegraphics[scale=0.8,angle=-00] {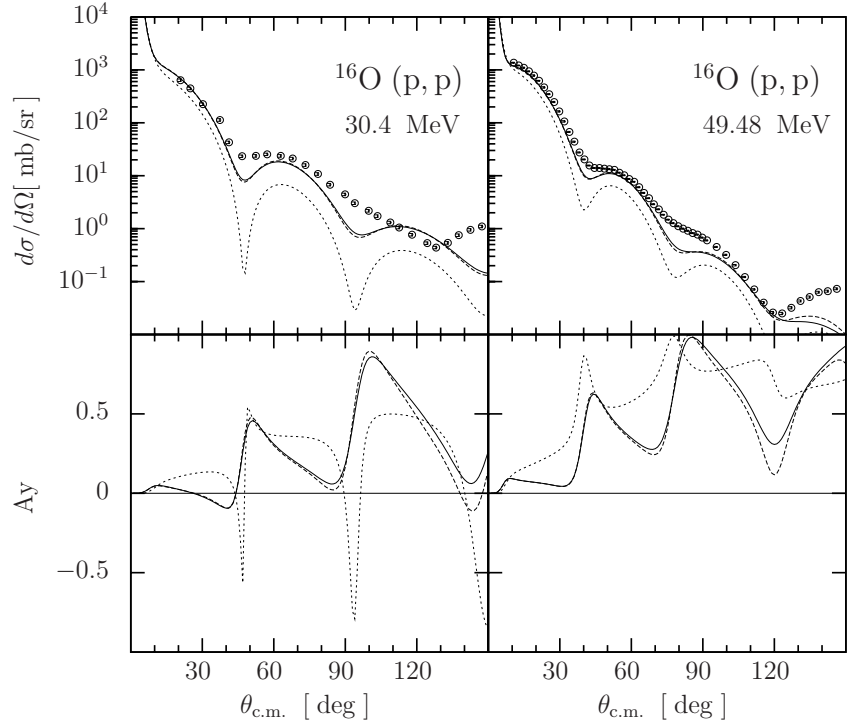}
\medskip
\caption{{\protect\small
\label{mca_O2} }
   Measured and calculated differential cross section for
$^{16}$O(\emph{p},\emph{p})  at 30.4 and 49.48 MeV (upper frames),
and calculated analyzing power (lower frames).
The data are from Refs. \cite{Gre72} and \cite{Fan67}, respectively.
See text for reference to the curves.
        }
\end{figure}

In Fig. (\ref{mca_O2}) we show the measured
and calculated differential cross section (upper frames) and
analyzing power (lower frames), as a functions
of the center-of-mass scattering angle, for $^{16}$O(\emph{p},\emph{p}) 
at 30.4 and 49.48 MeV.  The cross section data are 
from Refs. \cite{Gre72} and \cite{Fan67}, respectively.
Here again we observe that the $\delta g$-folding and ABL approach
follow very close each other.
Also, the free $t$ matrix results agree poorly with the data, as expected.
This lighter target evidences a deviation of the
calculated cross sections with the data, particularly the
depth of the diffractive minima in the cross section. 
Indeed, both $\delta g$- and ABL-folding approaches fail to account 
for the shallow first minima near 40$^\circ$.
In turn, both yield non existing minima near 90$^\circ$ and 80$^\circ$, 
respectively.

We have performed various tests of sensitivity to assess the
consistency of the results presented here. 
For instance, using harmonic-oscillator wavefunctions
for $^{16}$O --with the same root-mean-squared radii--  
we obtain practically the same results for the scattering observables. 
To keep this work focused on the structure of the
$\delta g$-folding, we have not explored the sensitivity of the
calculated scattering to alternative representations of the
target ground state, leaving such study for future works.

\begin{figure}[!ht]
\includegraphics[width=0.8\textwidth]{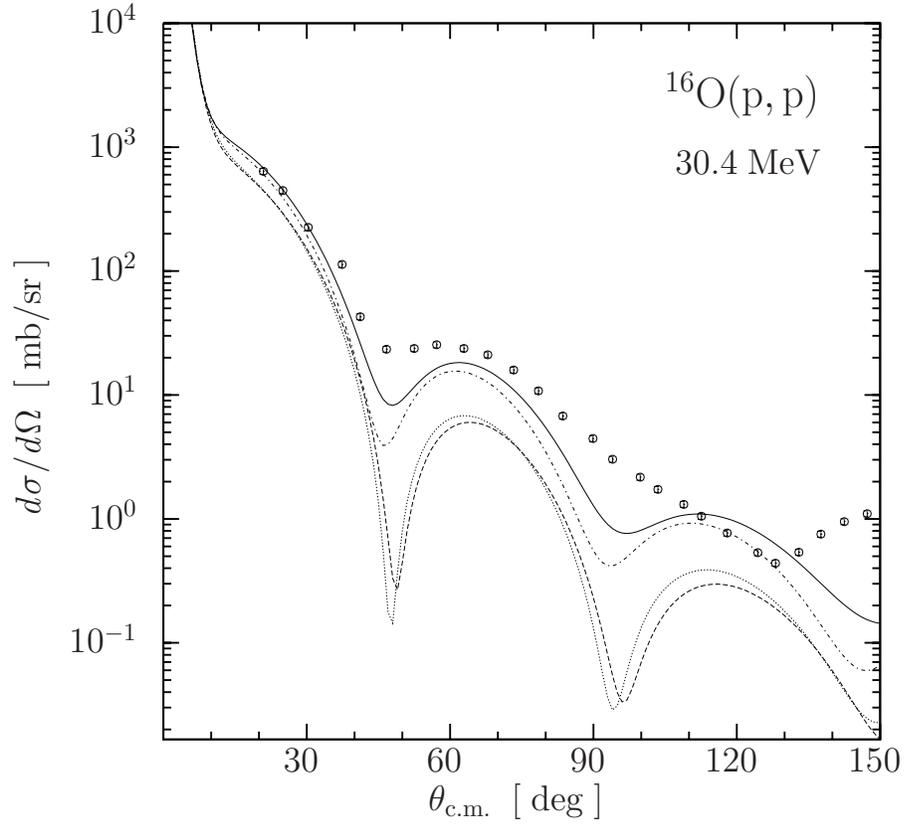}
\caption{
The calculated differential cross section based on the 
$\delta g$-folding (solid curve) and $t$-folding
(dotted curve) for $^{16}$O(\emph{p,p}) scattering at 30.4 MeV.
The data are from \cite{Gre72}.
See text for reference to the dashed curves.
 }
\label{mca_fit}
\end{figure}

In Fig. (\ref{mca_fit}) we examine how the differential cross
section is affected upon changes on $U_1$. 
The case is $^{16}$O(\emph{p},\emph{p}) at 30.4 MeV.
Here the solid curve represents results from the $\delta g$-folding,
$U_0+U_1$, while the dotted curve is based on $U_0$ alone,
namely the $t$-folding.
What is interesting to note is the effect of suppressing selectively the 
proton density (dash-dotted curve) and
neutron density contributions to $U_1$ (dashed curve).
We note that the role of $U_{1}^{(n)}$ is considerably more 
significant than that of $U_{1}^{(p)}$.
Indeed, by neglecting $U_{1}^{(n)}$ the cross section ends up
being very similar to the one obtained with the $t$-folding, 
in contrast with the moderate change on the $\delta g$-folding 
result when $U_1^{(p)}$ is suppressed.
These results are consistent with our analysis of $u_\alpha(Z)$
discussed in the previous section, confirming the importance of
neutron distributions in the optical model potential at these low energies.

\section{Summary and conclusions}
We have investigated the structure of the optical model potential
as inferred from its general form, once the MCA
is applied to the vector structure of the \emph{NN} couplings. 
The resulting ($\delta g$-folding) potential becomes expressed 
as the sum of two components, $U_0+U_1$, 
where $U_0$ corresponds to the free $t$ matrix full-folding potential 
and $U_1$ folds the full mixed density with the gradient of the 
medium-dependent effective interaction.
This feature implies that the intrinsic medium effects are
localized mainly in the nuclear surface.
When comparing the relative strength of these contributions, 
we find that the \emph{pn} coupling is considerably stronger than
its \emph{pp} counterpart, a feature that fades out as the energy of 
the projectile is increased. 
This asymmetry leads to stronger medium-sensitivity of proton 
scattering to neutron matter distributions of the nucleus.

As a by product of this study,
with the introduction of the $\delta g$-folding we have been able 
to provide a thorough account of the full mixed density in the evaluation
of momentum-space optical potentials within the $g$ matrix.
With this we mend a long standing limitation of the ABL folding 
approach to \emph{NA} scattering, where the mixed density has been
approximated by its Slater form.
When comparing the differential cross section,
the $\delta g$-folding and ABL approach are close to one another 
within 10\% in the diffractive maxima.

We have assessed the predicting power of the $\delta g$-folding 
approach to proton elastic scattering from $^{16}$O and $^{90}$Zr
at energies between 30 and 100 MeV. 
In the case of $^{90}$Zr(\emph{p},\emph{p}) we are able to provide
reasonably good  descriptions of the data.
For $^{16}$O(\emph{p},\emph{p}), in turn, the differential cross section 
is underestimated significantly. 
At this point we are not clear whether these shortcomings
stem from missing contributions implied when the dependence 
of $\partial g/\partial Z$ on ${\bm Q}$ is neglected,
the existence of exotic neutron structures in the surface,
or from  higher-order effects in the \emph{in-medium} effective interaction.
One has to keep in mind that the low-energy interaction
of the projectile with target nucleons becomes more sensitive
to the shell structure of the nucleus, in addition to the presence
of collective excitations or other reaction channels.
Investigations along these lines have recently been reported \cite{Fra08,Kee08}.

The present work constitutes a step forward toward a comprehensive
momentum-space description of the optical model potential for \emph{NA}
scattering, in the form of a unified description, for elastic
and inelastic processes from few tens of MeV up to GeV energies.
The only microscopic inputs to achieve this goal are the bare \emph{NN}
potential and the target ground-state mixed density,
although high-energy applications may also require \emph{NN}
phase-shift analyzes accounting for loss of flux above
pion-production threshold \cite{Are02}.
The introduction of the $\delta g$-folding optical potential has allowed us
to visualize very simply the interplay among different elements 
in the interaction of a single nucleon with finite nuclei,
particularly the role of the \emph{in medium} interaction
in the nuclear surface.
With the application reported here we have been able to set 
narrower margins of uncertainty in the evaluation of the 
first-order optical model potential, an important consideration 
for high-precision analyzes of upcoming scattering data involving 
unstable nuclei.


\begin{acknowledgments}
{\small H.F.A.} 
acknowledges partial support provided by 
{\small VID-UCH} under grant ENL0704, and 
{\small FONDECYT} under grant No 1080471.
\end{acknowledgments}


\end{document}